%% file: tng50mfs_avi.tex
\documentclass[twocolumn]{aastex631}
\usepackage{color}
\usepackage{xcolor}

\colorlet{avi}{green!0!orange!100!}

\definecolor{mhi}{rgb}{0.6,0.0,0.6}

\usepackage{float}
\usepackage{amsmath}

\newcommand{\NH}{N$_{\rm HI} $}
\newcommand{\cm}{$\rm cm^{-2}$}
\newcommand\pow[1]{\ensuremath{10^{#1}}}

\graphicspath{{./}{}}

\begin{document}

\title{Observational predictions for the survival of atomic hydrogen in simulated Fornax-like galaxy clusters}

\author[0000-0002-4175-4728]{Avinash Chaturvedi}\email{avi.chaturvedi@aip.de}
\affiliation{Leibniz-Institut f{\"u}r Astrophysik Potsdam (AIP), An der Sternwarte 16, D-14482 Potsdam, Germany}

\affiliation{European Southern Observatory, Karl-Schwarzschild-Stra\ss{}e 2,   ,85748 Garching, Germany}

\author[0000-0002-8710-9206]{Stephanie Tonnesen}
\affiliation{Center for Computational Astrophysics, Flatiron Institute, 162 5th Ave, New York, NY 10010, USA}

\author[0000-0003-2630-9228]{Greg L. Bryan}
\affiliation{Department of Astronomy, Columbia University, 550 W 120th Street, New York, NY 10027, USA}
\affiliation{Center for Computational Astrophysics, Flatiron Institute, 162 5th Ave, New York, NY 10010, USA}

\author[0000-0003-1151-4659]{Gerg\"o Popping}
\affiliation{European Southern Observatory, Karl-Schwarzschild-Stra\ss{}e 2,
85748 Garching, Germany}

\author[0000-0002-2363-5522]{Michael Hilker}
\affiliation{European Southern Observatory, Karl-Schwarzschild-Stra\ss{}e 2,
85748 Garching, Germany}

\author[0000-0001-5965-252X]{Paolo Serra}
\affiliation{INAF – Osservatorio Astronomico di Cagliari, Via della Scienza 5, 09047 Selargius (CA), Italy}

\author[0000-0002-3185-1540]{Shy Genel}
\affiliation{Center for Computational Astrophysics, Flatiron Institute, 162 5th Ave, New York, NY 10010, USA}

\begin{abstract}

The presence of dense, neutral hydrogen clouds in the hot, diffuse intra-group and intra-cluster medium is an important clue to the physical processes controlling the survival of cold gas and sheds light on cosmological baryon flows in massive halos. Advances in numerical modeling and observational surveys means that theory and observational comparisons are now possible. In this paper, we use the high-resolution TNG50 cosmological simulation to study the HI distribution in seven halos with masses similar to the Fornax galaxy cluster. Adopting observational sensitivities similar to the MeerKAT Fornax Survey (MFS), an ongoing HI survey that will probe to column densities of $10^{18}$ cm$^{-2}$, we find that Fornax-like TNG50 halos have an extended distribution of neutral hydrogen clouds. Within one virial radius, we predict the MFS will observe a total HI covering fraction around $\sim$ 12\% (mean value) for 10 kpc pixels and 6\% for 2 kpc pixels. If we restrict this to gas more than 10 half-mass radii from galaxies, the mean values only decrease mildly, to 10\% (4\%) for 10 (2) kpc pixels (albeit with significant halo-to-halo spread).  Although there are large amounts of HI outside of galaxies, the gas seems to be associated with satellites, judging both by the visual inspection of projections and by comparison of the line of sight velocities of galaxies and intracluster HI.

\end{abstract}


\textbf{\section{Introduction}\label{sec:intro}}

The formation and evolution of galaxies is now understood to be strongly linked with the diffuse gas filling their dark matter halos. 
Depending on galaxy mass, this gas is called the circumgalactic medium (CGM) when the gas resides in the halos of galaxies of the Milky Way mass or lower, or the intracluster medium (ICM) for gas living in more massive cluster-sized halos ($\rm M_{tot}$ $\gtrsim$ 10$^{14}$ M$_{\odot}$). 
Various physical processes, such as gas accretion from the intergalactic medium, and feedback driven by stars and AGN, occur in this gaseous halo. 
These processes drive flows which regulate the rate at which gas cools on to the galaxy itself, controlling the amount of mass in the interstellar medium and hence the rate of star formation itself, including possibly quenching \citep[see][for a review]{tumlinson2017}.
Therefore, understanding the physical processes in the CGM or ICM is crucial to building a comprehensive picture of galaxy evolution. We highlight here the important question of the origin and survival of cold gas in the hot ICM, by studying the phase-space distribution of cold gas in massive halos at low redshift, with a particular focus on the observability of this gas in ongoing and future HI surveys.

For low-mass, star-forming galaxies, it is expected that their halo can host an abundant amount of cold gas that may be the fuel for future star formation. The halos of massive galaxies, on the other hand, may lack cold gas, resulting in the lack of any recent star formation activity \citep{Gauthier2011}. In addition, current observational \citep{chen2018, Berg2019, zahedy2019} and simulation studies \citep{dave2020, rahmati2015, nelson2020} suggest that at intermediate redshift (0.3 $\leq$ z $\leq$ 0.8), massive halos have a significant amount of cold gas. For example, \cite{chen2018, zahedy2019} studied luminous red galaxies (LRG) at redshift z $\sim$ 0.21-0.55 and found that these galaxies host a high column density of the cold gas tracers HI and MgII. Other similar studies point toward the same conclusion that LRGs host an abundant amount of cold gas \citep{zhu2014, lan2018, anand2021, anand2022}. At low-redshift, simulation studies of Milky Way-like galaxies show that their CGM also host substantial amounts of cold gas \citep{voort2019}.

Recently, simulations have made improved predictions for the cold gas distribution in halos that reproduce many of the observed galactic properties. For example, \cite{nelson2020} have shown that cold gas in the halos of LRGs can be attributed to the thermal instability triggered by local density perturbations. They suggest that these perturbations are related to gas stripped from infalling galaxies via tidal interactions, or ram pressure stripping. Performing a comparative study between cosmological and idealized simulations (individual galaxy halo simulations), \cite{Fielding2020} have also shown that cold gas extends to the virial radius for Milky Way-mass halos. They also suggest that non-spherical accretion and satellite galaxies contribute to the cold gas content in the outer halos. Previously, \cite{Navarro2018} performed a detailed study using the Illustris TNG100 simulation, investigating the HI abundance and clustering properties in halos for z $\leq$ 5. They showed that HI density profiles are sensitive to various processes such as AGN feedback and tidal stripping. For massive halos, they found that HI is mostly concentrated in their satellite galaxies, whereas for small halos it is concentrated in the central galaxy.

In contrast to intermediate redshifts, the study of cold gas in the ICM of massive halos ($\rm M_{*}$ $\geq 10^{11}$ $\rm M_{\odot}$) in the local Universe (z $\sim$ 0) is limited to a few studies. Nonetheless, using the HI 21 cm emission line, radio observations have demonstrated the abundant existence of cold neutral atomic gas around early-type galaxies (E/S0) galaxies \citep{serra2012, serra2013, young2014, serra_atca_2013}. However, these observations are limited to tens of kpc around the targeted galaxies and are typically not sensitive to HI column densities below $\leq$ $10^{19}$ $\rm cm^{-2}$, and therefore do not provide a comprehensive picture of the cold gas in the ICM. 

The ongoing observations from the MeerKAT Fornax survey \citep[hereafter MFS,][]{serra2019, MFS2023}, a radio continuum and line survey of the Fornax cluster, provides an excellent opportunity to study the HI gas in great detail in the nearby Fornax ICM (d $\sim$ 20 Mpc). Photometric and spectroscopic studies \citep{cantiello2020, chaturvedi2022} have shown that the Fornax cluster mass assembly is still ongoing, making it an interesting target to study. MFS is dedicated to studying the HI distribution and kinematics within the Fornax environment. The HI column density sensitivity of the MFS ranges from $\sim$5$\times 10^{19}$~cm$^{-2}$ at a spatial resolution of $\sim$10 arcsec ($\sim$1 kpc at Fornax distance) down to $10^{18}$~cm$^{-2}$ at $\sim$100 arcsec ($\sim$10 kpc at Fornax distance). With a mosaic area of the 12 square degrees, MFS will detect HI in the Fornax intracluster (hereafter IC) region - which in this paper refers to the region within the massive dark matter halo and outside satellite galaxies.

The high-resolution TNG50 cosmological simulations \citep{nelson2019, pillepich2019} provide a median spatial resolution of $\sim$100 parsec and its validation of cold gas (neutral and molecular hydrogen) against observational work \citep{popping2019, Diemer2019} makes TNG50 an ideal framework to explore the cold gas distribution in Fornax-like halos. This also provides a chance to forecast the upcoming MFS survey results and test the simulations against the MFS observations. In this work, we use the TNG50 simulations \citep{nelson2019, pillepich2019} and adopt the observing criteria of the MFS to study the HI content in the TNG50 halos similar to the Fornax galaxy cluster. We also study the HI distribution in these halos and their IC region. We calculate the HI covering fraction for these halos and predict the expected observed MFS HI covering fraction. 

In addition, both the spatial and velocity distribution of HI gas in the ICM of clusters and groups can be used to gain insight into the origin and survival of this cold gas. If the gas is correlated in both position and velocity with satellites, we can argue that the HI is likely either stripped from satellites or cooling is induced by satellites.  However, if cold gas is not correlated with satellite galaxy positions or velocities, we might argue that either cold gas formation is related to the central galaxy or that cold gas survives in the ICM long enough to become virialized \citep[e.g.,][]{Voit2017, Rohr2023}.  In this paper we take the first step of making these spatial and velocity maps, leaving gas particle tracking to future work.

The paper is organized as follows: In section \ref{sec:analysis}, we briefly introduce the TNG50 simulation and present the methodology for calculating the HI covering fraction. Section \ref{sec:results} presents our results about the HI distribution and its covering fraction. In section \ref{sec:discussion}, we compare our results to current simulation (Section \ref{sec:dis_sims}) and observational (Section \ref{sec:dis_obs}) studies, and discuss the likely origin of the cold gas in the intracluster medium of Fornax-like halos (Section \ref{dis:origin}). Section \ref{sec:summary} presents the summary of the work. \\

\section{Simulation and Methodology}\label{sec:analysis}

This section briefly introduces the TNG50 simulation that we use for our analysis as well as our criteria for selecting Fornax-like halos. In addition, we present our methodology for calculating the HI column density and HI covering fraction of the selected halos in TNG50 in order to compare to observational surveys. 

\subsection{The TNG simulations}

For our study, we use the TNG50 simulation \citep{nelson2019, pillepich2019}, the highest resolution simulation of the IllustrisTNG cosmological magneto-hydrodynamical (MHD) simulation suite \citep{Nelson2018, Springel2018, Pillepich2018, Marinacci2018, Naiman2018}. The IllustrisTNG project is a set of large cosmological simulations that include a variety of galaxy formation physics including AGN feedback. The model has been designed to match a wide range of observational constraints \citep{Pillepich2018, Springel2018} and was carried out with the moving mesh code AREPO  \citep{springel2010}. The AREPO code solves the coupled evolution of dark matter, gas, stars, and black holes under the influence of self-gravity and ideal MHD. Developed with the key motivation to study galaxy formation physics and understand the growth of cosmic structure physics, the IllustrisTNG project uses three distinct simulation box sizes.

TNG50 was carried out with a box size of 51.7 Mpc per side with $2160^{3}$ gas and dark matter cells, resulting in a baryon mass resolution of $8.4\times10^{4}$ $\rm M_{\odot}$. In particular, we used TNG50-1 (hereafter TNG50), the highest resolution of the three variants run, which provides a median spatial resolution of $\sim$100 pc. We analyze this run, although the other larger boxes (TNG100 and TNG300) contain a larger number of Fornax cluster-sized objects, because we need high spatial resolution to study the interaction of cold high density columns and the hot group medium. TNG50 adopts initial conditions and cosmological parameters consistent with the \cite{planck2016} cosmology with $h = 0.68$, $\Omega_{b}$ = 0.05, $\Omega_{m}$ = 0.31, $\Omega_{\lambda}$ = 0.69, and $\sigma_{8}$ = 0.82 and assuming a flat universe governed by a $\Lambda$ cold dark matter ($\Lambda$CDM) cosmology.

\begin{deluxetable*}{cccccc}
\tablenum{1}
\tablecaption{TNG50 halos similar to the Fornax galaxy cluster\label{tab1:tng50halo}}
\tablewidth{0pt}
\tablehead{\colhead{TNG50 Halo ID} & \colhead{Virial Mass ($M_{200}$)} & \colhead{Virial radius ($R_{200}$)} & \colhead{Total HI mass} & \colhead{Central galaxy vel. dispersion}  & \colhead{Halo members vel. dispersion}  \\
& \colhead{Log M$_{\odot}$} & \colhead{($\times $ 100 kpc)} & \colhead{Log M$_{\odot}$} & \colhead{km/s} & \colhead{km/s} }
\startdata                              
\input{tng50_fnal_halos.tex}
\enddata
\end{deluxetable*}

\subsection{Fornax-like Halo selection} 

In TNG50 a galaxy cluster and groups of galaxies are referred to as halo or FOF (hereafter referred as halo), identified through the friends-of-friends algorithm \citep{Davis1985}. Within each halo, the SUBFIND algorithm \citep{Springel2001} identifies the subhalos including the primary (central) galaxy and other satellite galaxies (hereafter referred as satellite). To find halos similar to the Fornax cluster in TNG50 at snapshot 99 (redshift z = 0), we applied a virial mass selection criterion analogous to the Fornax cluster mass \cite[$\rm M_{200}$ $\sim$ $5\times10^{13}$ $\rm M_{\odot}$, adopted from][]{Drinkwater2001}, namely the mass range of $10^{13.5}$ $\leq$ $\rm M_{200}$ $\leq$ $10^{14}$ $M_{\odot}$, where $\rm M_{200}$ is defined as the mass enclosed within a virial radius $\rm R_{200}$ equal to 200 times the critical density of the Universe. With this condition, we find a total of seven halos. 

For these halos, we measured the stellar velocity dispersion of their central galaxy and found that this value is quite close to that of NGC1399, the central galaxy of the Fornax cluster. Except for halo IDs 4 and 9, all other halo central galaxy stellar velocity dispersions fall within $\pm$ 20 \% to the stellar velocity dispersion value of 315 kms of NGC1399 (Vaughan 2019). In addition to this, the velocity dispersion of all the subhalos within these halos agrees to within 15\% of the observed Fornax cluster members (giants and dwarf galaxies) mean velocity dispersion value of 374 $\pm$25 $\rm km/s$ \citep{Drinkwater2001}.  A previous study of the hot X-ray emitting medium of galaxy groups in the TNG50 cosmological simulations \citep{Truong2020} showed a good match to observations. More recently, a new set of zoom-in cosmological simulations using the same model `TNG-cluster' \citep{Truong2023} have demonstrated good agreement on a larger sample. These successes in reproducing observable properties of galaxy clusters indicates that our seven halos are reasonable matches to the Fornax cluster.
From here onward, we refer to these halos as Fornax-like halos. In Table~\ref{tab1:tng50halo}, we list the physical properties of these halos. 


\subsection{Atomic HI content}
To determine the HI mass of gas cells in TNG50 Fornax-like halos, we use the \cite{popping2019} molecular hydrogen fraction ($\rm H_{2}$) catalogue, previously calculated for the TNG simulations.  In this work we use their fiducial recipe, which is based on the work by \citet{Gnedin2011}. 

\citet{Gnedin2011} performed detailed simulations including non-equilibrium chemistry and simplified 3D on-the-fly radiative transfer calculations. Based on these simulations, the authors presented fitting formulae for the $\rm H_{2}$ fraction of neutral gas as a function of the dust-to-gas ratio of the gas, the impinging UV radiation field, and surface density of the neutral gas. \citet{popping2019} assume that the dust-to-gas ratio scales with the metallicity of the neutral gas, that the local UV radiation field scales with the SFR of the gas cell with an additional contribution from the ionising UV background field and they calculate the gas surface density of a gas cell by multiplying its density by the Jeans length of the cell. A detailed description of the implementation of the \citet{Gnedin2011} fitting formulae within the TNG simulation suite is presented in \citealp[][(see their Section 2)]{popping2019}.

\begin{figure*}
\epsscale{1.15}
\plotone{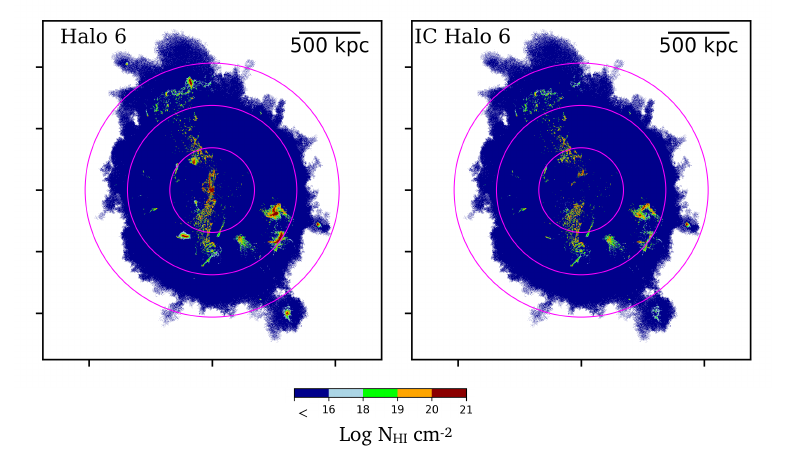}
\caption{HI distribution in the TNG50 halo 6 at redshift z=0, colour-coded with the HI column density. The maps are made using a pixel scale of 2 kpc and are shown projected along the (arbitrarily chosen) z-axis. The three pink circles indicate the viral radius of the halos marked at 0.5, 1.0 and 1.5 times $\rm R_{\rm vir}$. Dark blue color in the maps indicate the HI column density lower than \NH= \pow{16} \cm. {\it Left:} Full halo HI distribution, {\it right:} HI distribution in the intracluster (IC) region (i.e. after removing HI within 10 stellar half-mass radii of all galaxies).}
\label{fig:HIzoominmap}
\end{figure*}

\subsection{Halo HI covering fraction} \label{mock_maps}

To understand and quantify the HI distribution in the halos, we measure their HI covering fraction in different column densities bins (hereafter denoted as \NH), adopting a range of \NH $>$ \pow{18}, \pow{19}, and \pow{20} \cm. We first measured the HI column density, by performing a two-dimensional binning of all HI gas cells along the three projected axis X, Y and Z (spatial position), regardless of the velocity space and adopted a pixel size of 2 kpc, similar to the MFS spatial resolution limit. This assumes that gas particle sizes are smaller than 2 kpc, which is generally true for the TNG50 halos gas cells. We checked that larger gas cells have negligible contribution to the HI mass and hence to HI column density. In Section \ref{sec:HIcov_IC}, we present the HI covering fraction results as a function of velocity. 

We measure the HI covering fraction in two ways, similar to \cite{rahmati2015}, as follows:

\textit{Cumulative HI covering fraction}, hereafter denoted as $f_{HI}(R)$, is defined as the fraction of surface area covered by the binned pixels having column density higher than a given N$_{\rm HI}$ value within a radius R divided by the total area of pixels within radius $R$. $f_{HI}(R)$ is expressed as:

\begin{equation}
f_{HI}(R) = \frac{\sum_{i=1}^{N} A_{N_{HI}} \vert_{R}}{\sum_{i=1}^{N} A \vert_{R}}
\end{equation}
Here $A_{N_{HI}}$ is the single pixel area with a column density equal or higher than the given \NH {} value and is summed over $N$ such pixels in a given area of radius $R$ divided by the total area of pixels in radius $R$. 

\textit{Differential HI covering fraction}, hereafter denoted as $f_{HI}(\Delta R)$, is defined similarly to be cumulative HI covering fraction, except here we consider only the covering fraction within the radial bin defined between radius $R_{j}$ and $R_{j+1}$ and $f_{HI}(\Delta R)$ is expressed as:

\begin{equation}
f_{HI}(\Delta R) = \frac{\sum_{i=1}^{N} A_{N_{HI}} \vert_{\Delta R}}{\sum_{i=1}^{N} A \vert_{\Delta R}}
\end{equation}

We also separately measure the total HI covering fraction and the HI covering fraction in the intracluster (IC) regions of the halos, that is, the regions well away from identified galaxies. For the IC measurement, we first select the satellite galaxies within a halo with a stellar mass $\geq$ $10^{8.5}$ $\rm M_{\odot}$ and having at least 10 gas cells (varying this to 1 gas particle has no effect on our results). Then, we remove the HI gas cells associated with these satellite (including the central galaxy) using the SUBFIND algorithm of TNG50. This removal of HI gas is done out to ten times the stellar half mass radius (denoted as $\rm R_{1/2*}$) of a satellite. In TNG50, the SUBFIND algorithm identifies all the gas cells that are gravitationally bound to a specific satellite. After removing the gas cells of a satellite identified with the SUBFIND procedure out to 10$\times \rm R_{1/2*}$, we measure the HI covering fraction in the same way as defined earlier. The $\rm R_{1/2*}$ radii of satellite vary from a few kpc to tens of kpc, and with our adopted radial limit (10$\times \rm R_{1/2*}$), we make sure that we exclude all gas cells that are within the domain of an individual galaxy. However, the IC measurement may contain the very extended tidal/stripped gas tails originating from the individual galaxies.


\section{Results}\label{sec:results}

\begin{figure*}[!]
\epsscale{1.0}
\plotone{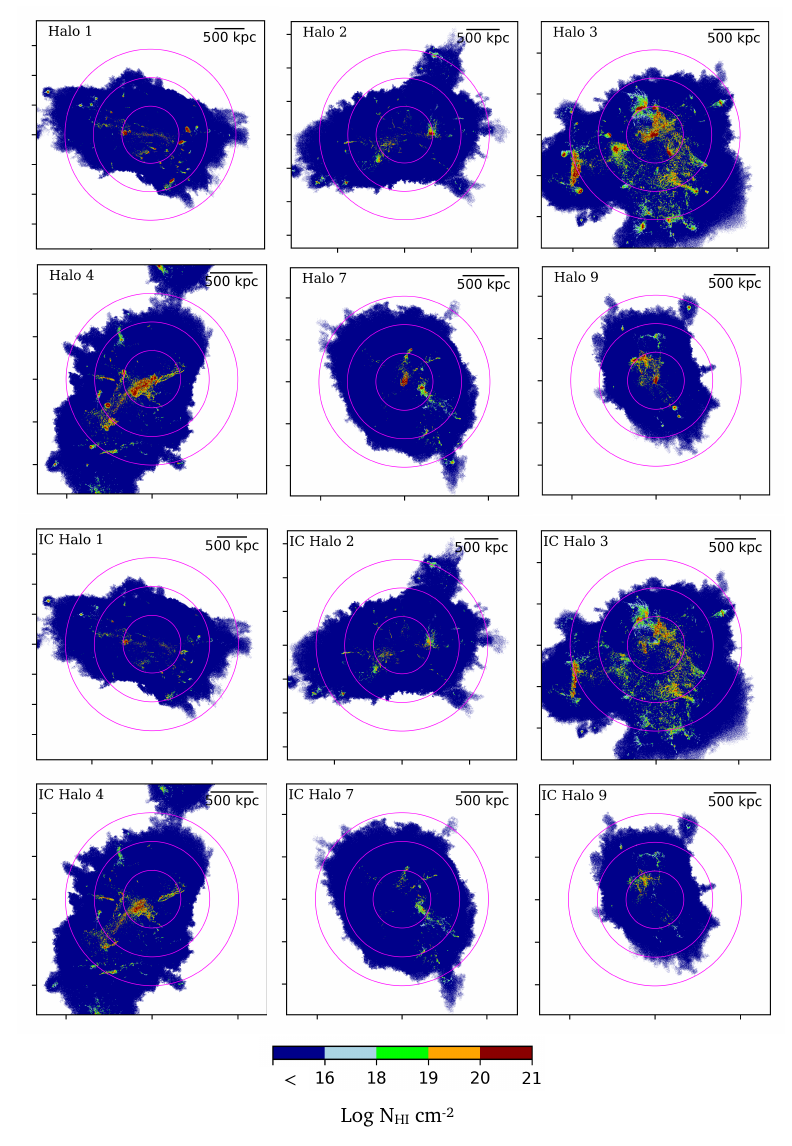}
\caption{HI maps in the Fornax-like halos (first two rows) and in their IC regions (bottom two rows) made with a pixel size of 2 kpc (projected along the arbitrarily chosen z-axis). The three pink circles indicate the viral radius of the halos marked at 0.5, 1.0 and 1.5 times $\rm R_{vir}$.}
\label{fig:HImaphalo}
\end{figure*}

\begin{figure*}[!]
\plotone{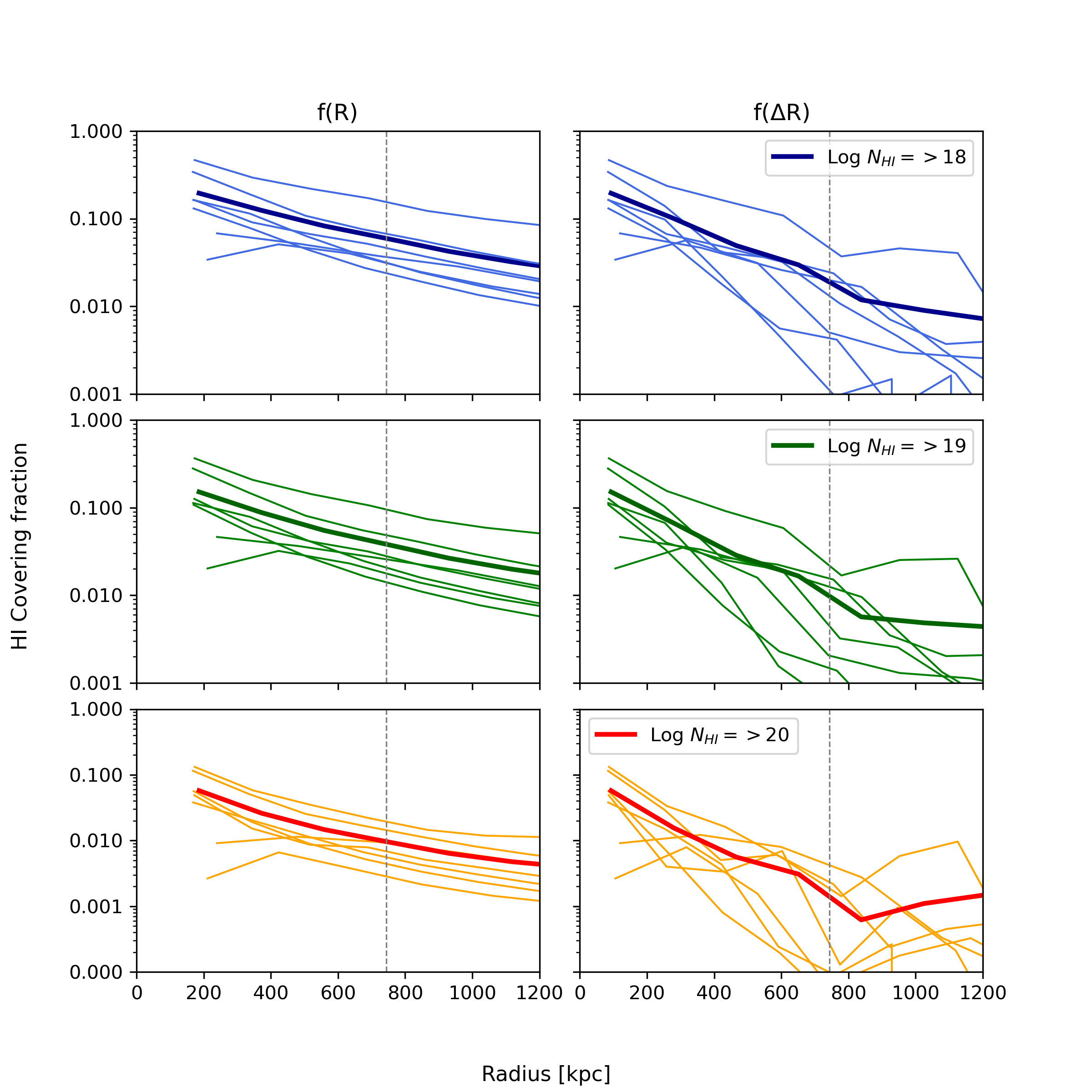}
\caption{HI cumulative (left panel) and differential (right panel) covering fraction profiles of Fornax-like halos HI maps (Figure \ref{fig:HImaphalo}, first two rows) measured along the arbitrarily chosen z-axis for a pixel size of 2 kpc. The first, middle, and bottom rows show the covering fraction for HI column densities log N$_{\rm HI}$ $\geq$ 18, 19 and 20 $\rm cm^{-2}$ respectively. The thin lines indicate the individual halos, and the thick lines mark the average value. The vertical dashed lines indicate the average virial radii of the halos.}
\label{fig:fcovhalo}
\end{figure*}

\begin{figure*}[!]
\plotone{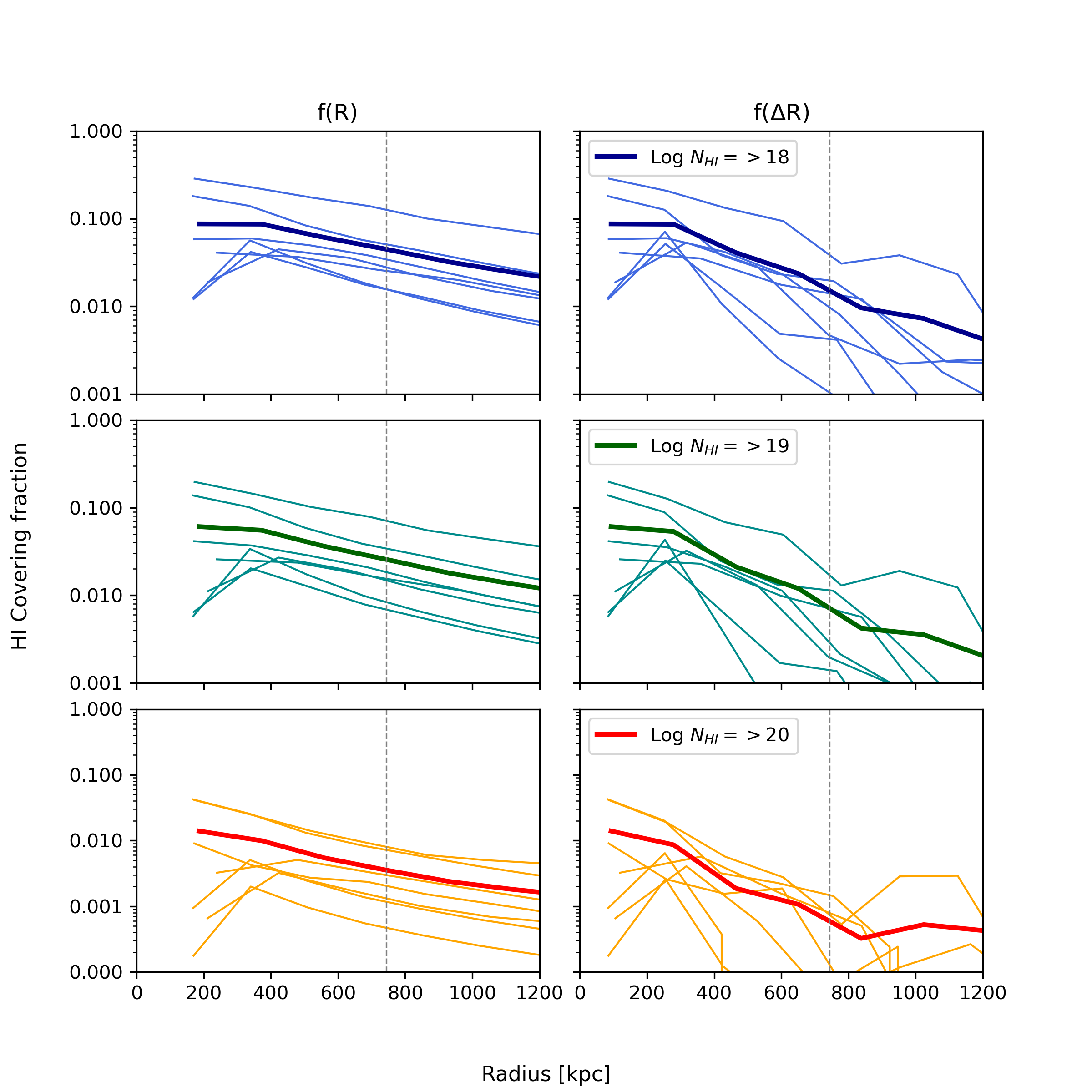}
\caption{The same as Figure \ref{fig:fcovhalo}, but for the HI distribution in the intracluster (IC) region of Fornax-like halos HI maps  (Figure \ref{fig:HImaphalo}, bottom two rows) measured along the arbitrarily chosen z-axis for a pixel size of 2 kpc. We obtain the IC HI by removing the gas cells gravitationally bound to all galaxies within 10$\times$ their stellar half mass radii.  Note the y-axis ranges differ from Figure \ref{fig:fcovhalo}.}
\label{fig:fcovIC}
\end{figure*}

This section presents our results showing the HI distribution and covering fraction in the TNG50 halos and their IC regions. For our study, we considered gas cells gravitationally bound to a halo, including gas cells extending to an average 1.5 $\rm R_{\rm vir}$ radii of halos. Figure~\ref{fig:HIzoominmap} shows the HI distribution in TNG50 halo 6 (left panel) and in its IC region (right panel). Three pink circles indicate the virial radius drawn at 0.5, 1.0 and 1.5 $\rm R_{\rm vir}$. These images demonstrates both the patchy nature of HI in the simulated clusters, as well as it's distributed nature. The IC image (right) emphasizes that much of the HI (at least by area) is not immediately connected to galaxies -- that is, it is at least 10 stellar half-mass radii from any galaxy in the simulation.

Figure \ref{fig:HImaphalo} shows the HI distributions for all other Fornax-like halos (top two rows) and in their IC regions (bottom two rows). The first visual impression we get from these plots is that the large-scale distribution of HI extends beyond 0.5 virial radii ($\sim$350 kpc) for all the halos. Otherwise, these halos demonstrate diverse HI distributions with significant variations in the amount of HI mass (Table \ref{tab1:tng50halo}). The diverse and extended HI distribution in these halos could potentially be related to the merger/accretion history of these halos or the activity of supermassive black holes \citep{Zinger2020}. We can also see the streams or filamentary structures connecting the central and satellite galaxies. {We notice, in addition, a large number of small HI regions with relatively high column densities, which are possibly not related to any satellite. We refer to these as clouds. 
We caution that, unlike \cite{nelson2019}, we have not performed Voronoi tessellation over the gas cells to identify these clouds-like structures and it merely represents HI clumps around the satellite galaxies and in the intra-cluster regions of these halos. For the halos HI map, we see that the centers of the satellite are dominated by HI column densities \NH\ between \pow{20} to \pow{21} \cm. Looking at these maps, it is quite clear that HI clouds extend out to the virial radius (corresponding to $\sim$ 700 kpc) covering the IC regions. Within the inner region of each halo, around $\sim$ 0.25 $\rm R_{\rm vir}$, a large fraction of the HI gas cells are associated with the central galaxy of the halo. In the IC region, the observed HI structures primarily have column densities \NH $<$ \pow{20} \cm, whereas the structures beyond 0.5 virial radii lie mainly at \pow{18} $<$ \NH/\cm $<$ \pow{19}. We present the results of the HI covering fraction of halos and in the IC regions in subsections \ref{sec:HIcov} and \ref{sec:HIcov_IC}, respectively. 

\subsection{HI covering fraction profiles}\label{sec:HIcov}

We used the HI projected maps as shown in Figure~\ref{fig:HIzoominmap} to measure the HI covering fraction as discussed in section~\ref{sec:analysis}. We measured the HI covering fractions of halos for three projections, along the X, Y, and Z directions, using a pixel scale of 2 kpc. 

In this section, we present the HI covering fraction for the full halo maps (top two rows in Figure \ref{fig:HImaphalo}) and in the next section, we discuss the covering fraction of IC regions. In Figure \ref{fig:fcovhalo} we show the cumulative and differential covering fraction profiles measured along the (arbitrarily chosen) z-axis in the left and right panel, respectively. For these panels, we include all of the HI gas in the covering fraction calculation, whether or not it is within the central or a satellite galaxy.  We do this because, in a blind HI survey, the satellite galaxies may not be identified, so the HI would be measured globally. By definition, the innermost point of the cumulative and differential covering fractions are the same, then the differential covering fraction begins decreasing more steeply than the cumulative covering fraction. The first, middle, and bottom rows in both panels show the covering fraction for HI column density for \NH\ $\geq$ \pow{18}, \pow{19} and \pow{20} \cm\ respectively. The thin lines indicate the individual halos, and the thick lines mark the average value.

We focus mainly on the \NH\ bins of \pow{18} and \pow{19} \cm, which are the optimal range for studying the HI distribution at $\sim$kpc scales for the MFS. We find that regardless of the projection axis, the average covering fraction for the \NH\ $\geq$ \pow{18} \cm\ bin remains between 10-15\% at 0.5 $\rm R_{\rm vir}$ and drops to 6-10\% at 1 $\rm R_{\rm vir}$. With increasing column density, the covering fraction decreases, such that for the 
N$_{\rm HI}$ $\geq$ \pow{19} \cm\ bin the covering fraction drops to 5-10\% at 0.5 $\rm R_{\rm vir}$ and to less then 5\% at 1 $\rm R_{\rm vir}$. The differential covering fraction at 0.5 virial radius is between 5-10\% and drops to less than 5\% at 1 $\rm R_{\rm vir}$.

These covering fractions quantify our visual impressions. Although all the halos have some HI gas within 0.5 $\rm R_{\rm vir}$, it is distributed non-uniformly in small structures that look like filaments or clouds.  In Figure \ref{fig:fcovhalo} we verify that the covering fraction of these structures is low, even when including column densities down to $10^{18}$ cm$^{-2}$. These structures and clouds could potentially be associated with the central galaxy and satellite galaxies, but from Figure \ref{fig:HImaphalo} we see that the HI is clearly extended well beyond the satellite stellar radii.  However, this gas might have been stripped from satellites to form part of the IC region, which we discuss in the next section and in Section \ref{dis:origin}. 

\begin{figure}
\epsscale{1.2}
\plotone{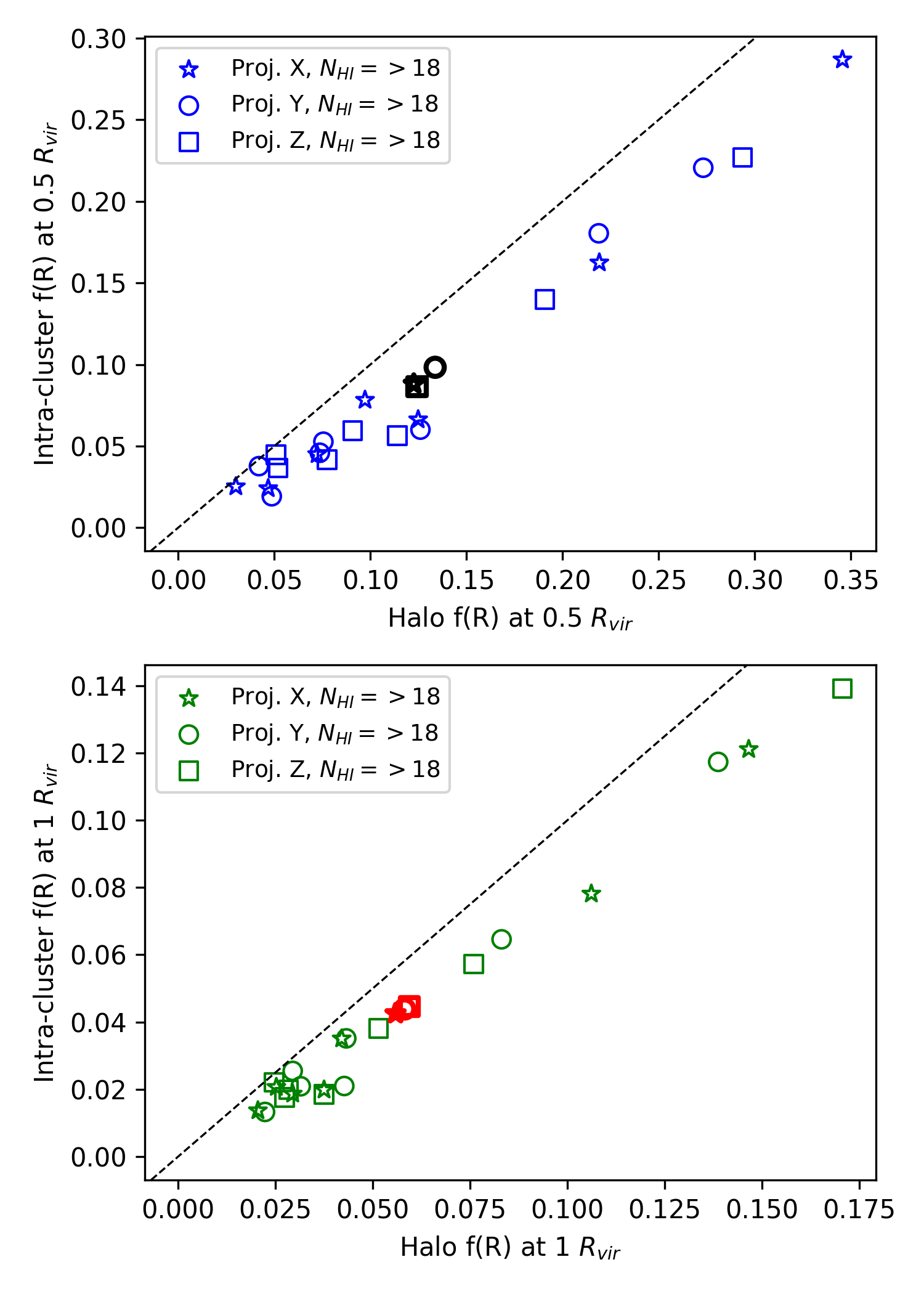}
\caption{Halo and intra-cluster cumulative HI covering fractions at 0.5 $\rm R_{vir}$ (upper panel) and 1 $\rm R_{vir}$ (lower panel) for three projected axes. Open star, circle and square indicate the X, Y and Z projections, respectively. The open black (red) star, circle and square indicate the average values.}
\label{fig:scatter}
\end{figure}

\begin{figure}
\epsscale{1.2}
\plotone{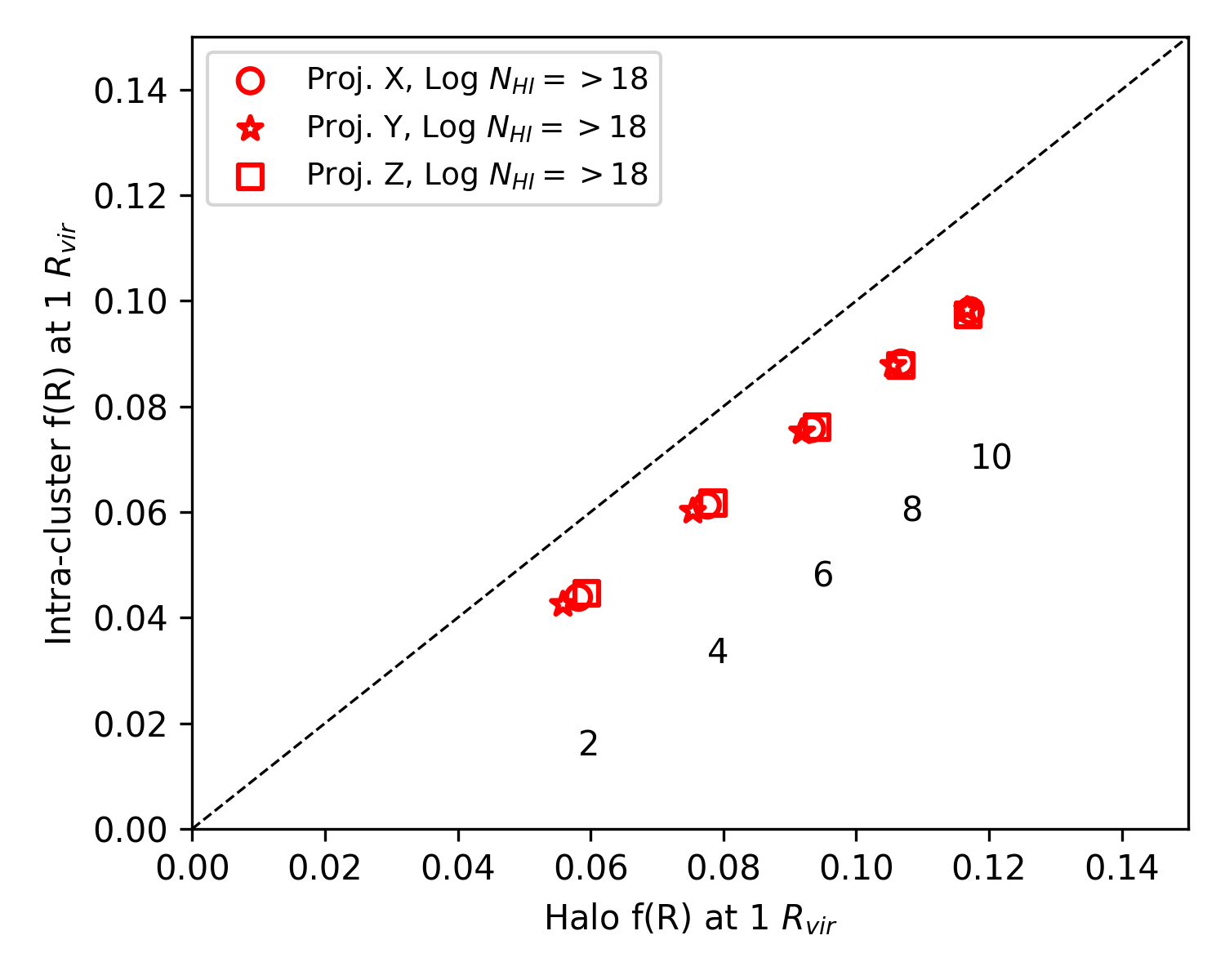}
\caption{Average halo and intra-cluster cumulative HI covering fractions of seven halos at 1 $\rm R_{vir}$ for three projected axes corresponding to different pixel scale size of HI maps. Numbers in the plot mark the pixel size in kpc used in creating the HI maps.}
\label{fig:cov_fraction_vs_pixelsize}
\end{figure}

\begin{figure*}[!]
\epsscale{1.15}
\plotone{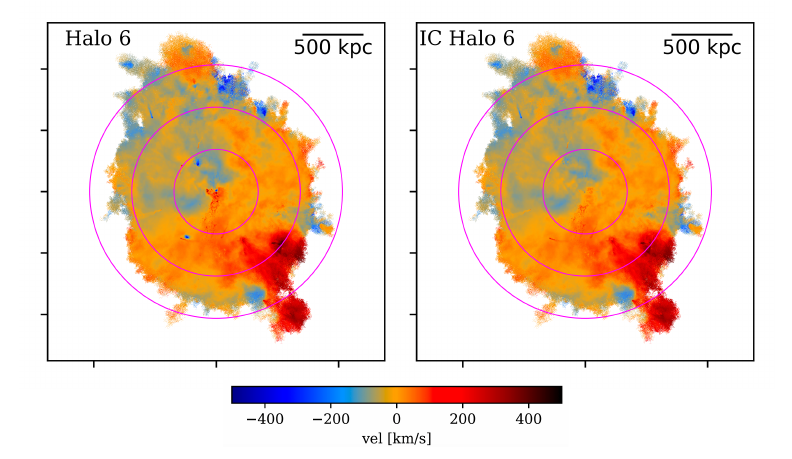}
\caption{HI distribution in the TNG50 halo 6 at redshift z=0, colour-coded with the HI mean velocity (projected along the arbitrarily chosen Z axis) for a pixel size of 2 kpc. \textbf{Left}: Full halo HI distribution, \textbf{right}: HI distribution in the IC region.}
\label{fig:HIzoominmap_vel}
\end{figure*}

\begin{figure*}[!]
\epsscale{1.00}
\plotone{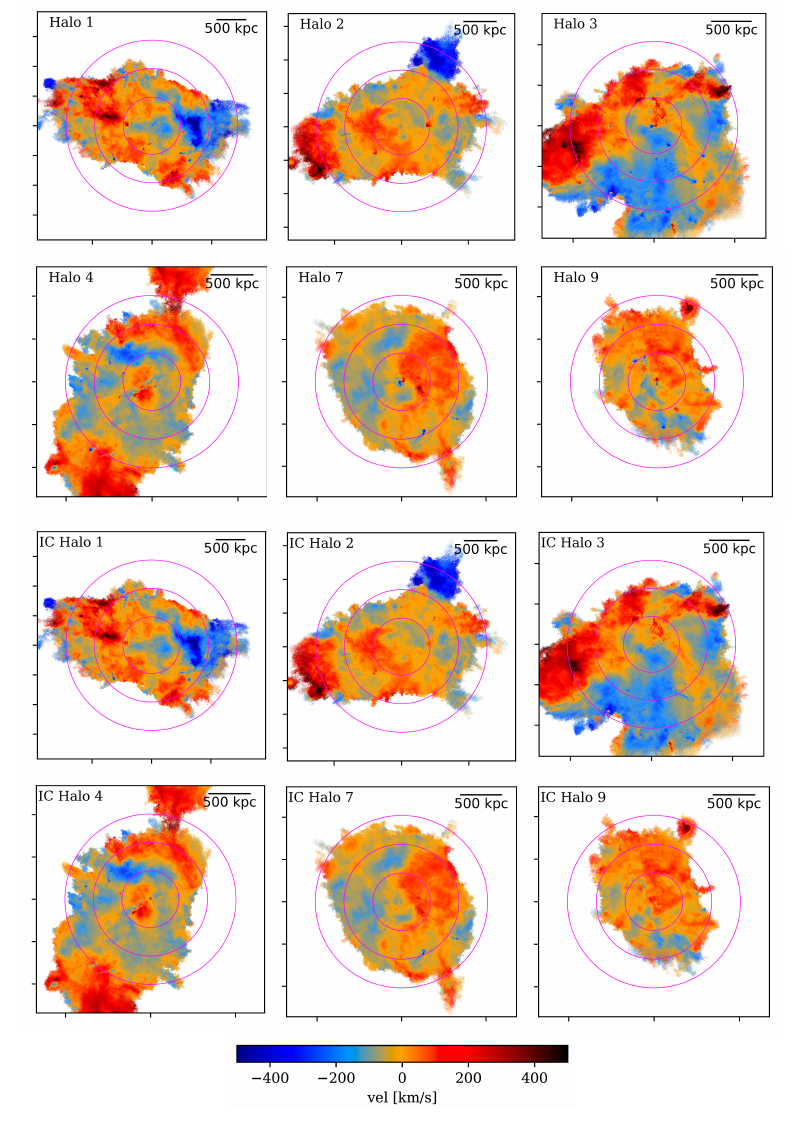}
\caption{HI distribution in the TNG50 halos (top two rows) and in their IC (bottom two rows) at redshift z=0, colour-coded with the HI velocity (projected along the arbitrarily chosen Z axis) for a pixel size of 2 kpc.}
\label{fig:velmaphalo}
\end{figure*}

\begin{figure}
\epsscale{1.15}
\plotone{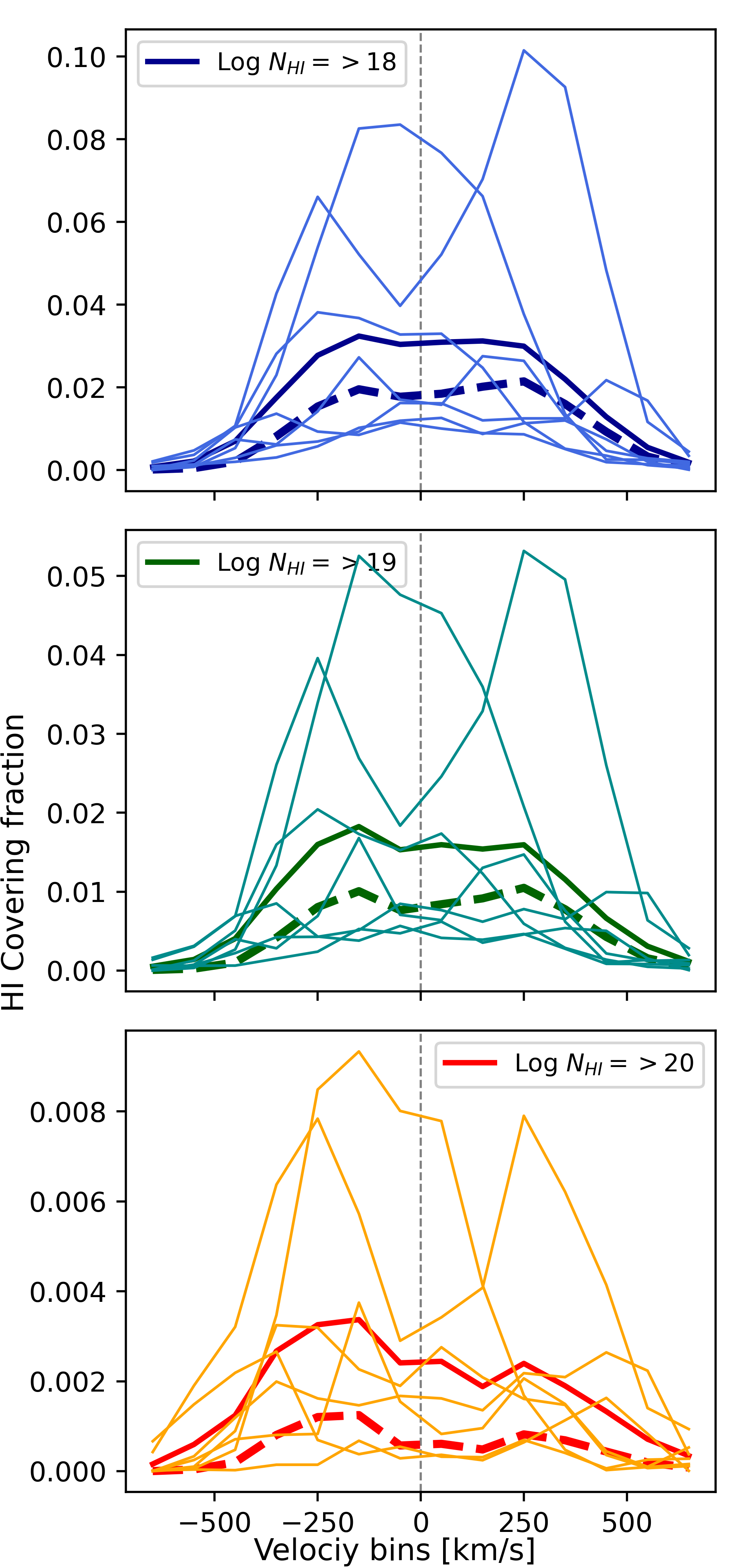}
\caption{HI cumulative  covering fraction profiles of Fornax-like halos in the velocity space. The top, middle, and bottom panels show the covering fraction for HI column densities log N$_{\rm HI}$ $\geq$ 18, 19 and 20 $\rm cm^{-2}$, respectively. The thin lines indicate the individual halos, and the thick lines mark the average value. The thick dashed lines mark the average IC HI cumulative covering fraction in velocity space.}
\label{fig:fcov_velocity}
\end{figure}

\begin{figure*}[!]
\epsscale{1.20}
\plotone{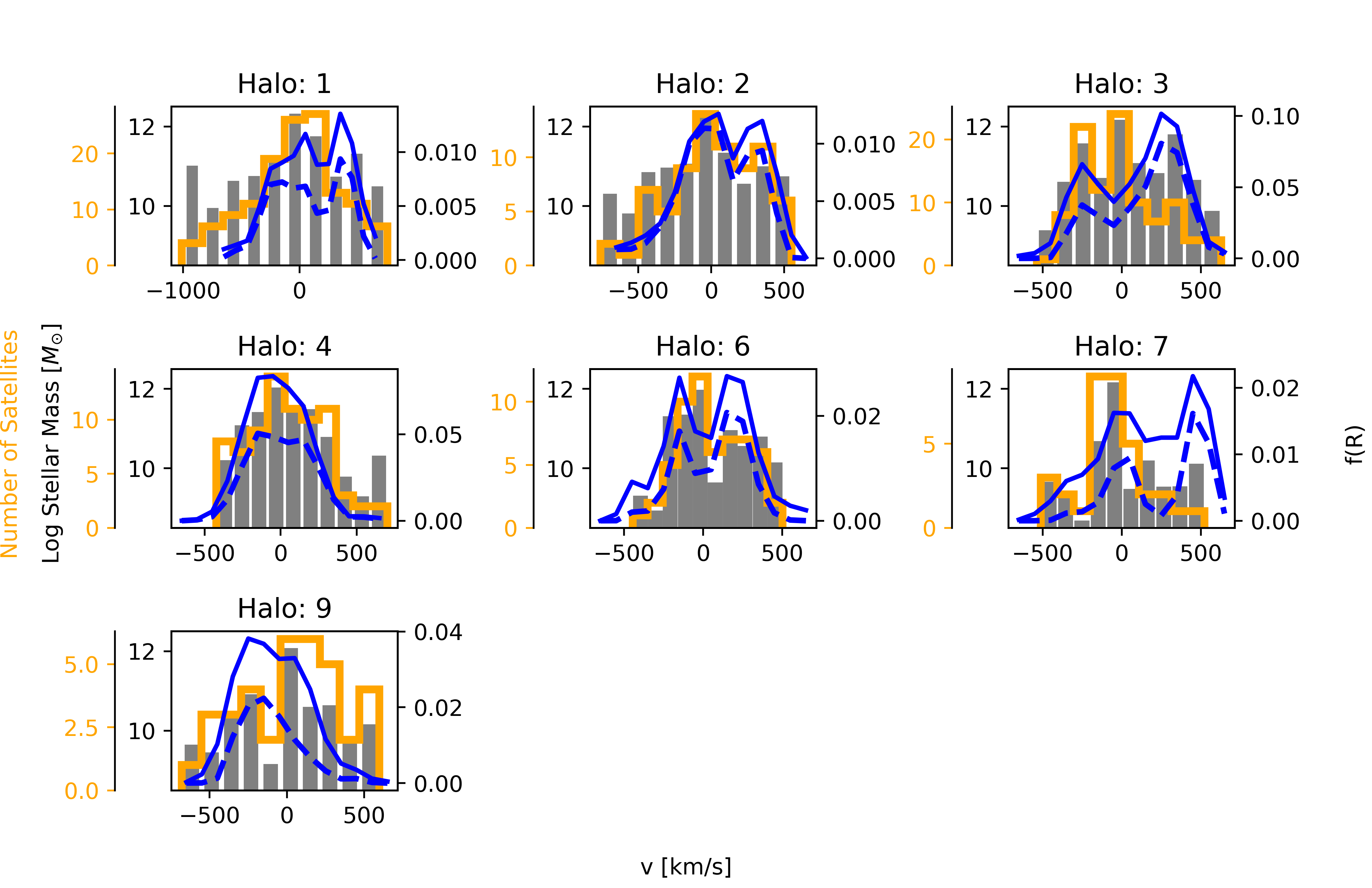}
\caption{HI covering fraction profiles of Fornax-like halos in velocity space for log N$_{\rm HI}$ $\geq$ 18 $\rm cm^{-2}$ gas over-plotted on the number counts of all satellite galaxies (orange histogram) and the stellar mass velocity distribution (grey histogram). The profiles and histograms are measured in 100 km/s bins. The leftmost y-axis (orange) of each subplot represents the number count of satellite galaxies, while the second axis (black) represents the log stellar mass of satellite galaxies. Solid and dashed blue lines indicate the velocity covering fraction of the halo and IC regions, respectively (right y-axis of each panel).}
\label{fig:fcov_velhist}
\end{figure*}

\subsection{Intra-cluster HI covering fraction} \label{sec:HIcov_IC}

The bottom two rows of Figure \ref{fig:HImaphalo} show the HI distribution in the IC regions of the halos. The majority of pixels with 
column density \NH $\geq$ \pow{18} \cm\ extend into the IC regions (the cumulative HI covering fraction drops by about 30\% at 1 R$_{vir}$). It is primarily pixels with high column density HI, \NH\ $>$ \pow{20} \cm, which are removed when only including the IC gas cells in our column density measurement (the cumulative HI covering fraction drops by about 70\% at 1 R$_{vir}$). 

Similar to Figure~\ref{fig:fcovhalo}, we show the HI covering profiles for the IC regions in Figure \ref{fig:fcovIC}. For the \NH\ bins \pow{18} and \pow{19} \cm, the cumulative covering fraction at 0.5  $\rm R_{\rm vir}$ is between 5-10 \% and drops to less than 5\% at 1 $\rm R_{\rm vir}$.  

We checked the differences in the HI covering fraction along the different projected directions and find that, on average, the HI covering fraction remains the same and varies only by a few percent when changing the projected axis. Figure \ref{fig:scatter} shows the cumulative HI covering fraction at 0.5 $\rm R_{vir}$ (upper panel) and 1 $\rm R_{\rm vir}$ (lower panel) between the Fornax-like halo and their IC region in three different projected axes for a column density \NH\ $\geq$ \pow{18} \cm. The open blue (green) star, circle and square indicate the covering fractions at 0.5 $\rm R_{\rm vir}$ (1 $\rm R_{\rm vir}$) along the X, Y and Z axis projections, respectively. The open black (red) star, circle and square indicate the average values measured at 0.5 $\rm R_{\rm vir}$ (1 $\rm R_{\rm vir}$). On average, the IC HI covering fraction is between 70-80\% of the total HI covering fraction. This suggests that a large fraction, around $\sim$ 75\% (by covering fraction) of the low-column density HI gas that is distributed throughout these massive halos is well outside the satellite galaxies. 
We also note that some pixels that had a high column density when including the satellite gas have lower column densities when only including IC gas.  This can be seen as some red pixels in the top panels of Figure \ref{fig:HImaphalo} turning into green pixels in the bottom panels.  

Depending on the HI column sensitivity, MFS will map the HI distribution in Fornax at different spatial resolutions, varying from 1 to 10 kpc. We created similar HI maps, as shown in Figure \ref{fig:HImaphalo}, with different pixel sizes ranging from 2 to 10 kpc. Examples of HI maps made with 10 kpc pixel scale are shown in Appendix \ref{appen1}. In Figure \ref{fig:cov_fraction_vs_pixelsize}, we show the average halo and IC cumulative HI covering fraction of the 7 halos at 1 $\rm R_{vir}$ radius for different pixel scales. As anticipated, when degrading the pixel size of HI maps, the HI covering fraction increases and, for \NH $\geq$ \pow{18} \cm, reaches 12\% at 10 kpc pixel scale for the full halo and is around 10\% for the IC halo map. These are then our resolution-matched predicted values for the \NH $\geq$ \pow{18} \cm covering fraction that will be observed by MFS.

Finally, in order to investigate whether we can see any signature of diffuse HI gas in the inter-galactic medium or cosmic filamentary structures around these mock Fornax-like halos, we made HI maps extending out to 3 $\rm R_{\rm vir}$, which we show in Appendix \ref{appen2}. Visually inspecting these maps, we do not find diffuse HI in the inter-galactic medium of halos. We notice that outside 1.5 $\rm R_{\rm vir}$, there are several infalling smaller satellite galaxies, particularly for Halos 1, 2 and 4, but they contribute a negligible amount in the HI covering fraction profile. In particular, we computed HI covering fractions for these maps and found similar results as for the maps shown in Figure~\ref{fig:HImaphalo}.

\subsection{HI covering fraction in velocity space}\label{sec:vel_covering_frac}

We also measured the cumulative HI covering fraction in the line-of-sight velocity space, adopting a velocity bin size of $\sim$100 $\rm kms^{-1}$ within a range of -700 to 700 $\rm kms^{-1}$. This allows us to connect the HI gas at different column densities in velocity space. A correlation of HI gas in velocity space with the satellite velocity distribution can give us hints about its possible stripping origin from satellite galaxies. Figure \ref{fig:HIzoominmap_vel} shows the HI distribution in velocity space for Halo 6 (left panel) and in its IC region (right panel). Similar to Figure \ref{fig:HIzoominmap}, we created these velocity space maps (Sec. \ref{mock_maps}) by performing two-dimensional binning in the phase space and taking the mean velocities in bins corresponding to the X, Y and Z axis, with a pixel scale of 2 kpc. In the velocity maps, we consider the halo velocity as the zero velocity point. 
Remaining HI velocity maps of other halos and their IC regions are shown in Figure \ref{fig:velmaphalo}. In these maps, we can see the distribution of larger satellite galaxies having different velocities then the surrounding lower density HI (first two rows in Figure \ref{fig:velmaphalo}). We used these maps to measure the HI covering fraction in velocity space for these halos which is shown in Figure \ref{fig:fcov_velocity}. The top, middle, and bottom panels show cumulative HI covering fraction for \NH \ column densities of \pow{18}, \pow{19} and \pow{20} \cm\ with a velocity bin size of 100 $\rm kms^{-1}$. The thin lines indicate the individual halos, and the thick continuous and dashed lines mark the average value of halo and IC region respectively. Vertical dashed gray line marks the halo velocity which is taken as the zero velocity point. 

We find that, for a velocity bin size of 100 $\rm kms^{-1}$, on average the cumulative HI covering fraction is less than 4 \% (2 \%) for the column density \NH \ $\geq$ \pow{18} (\pow{19}) \cm\ and significantly drops to less than 1\% for \NH\ $\geq$ \pow{20} \cm. The halo and IC HI velocity covering fraction looks bimodal for most of the halos (except halo 4 and 9), as also shown by the average HI covering fraction (continuous and dashed blue lines in Figure \ref{fig:fcov_velhist}). On a halo-by-halo comparison, we have verified that the IC HI does not show a more Gaussian velocity distribution than the total HI, as would be expected by virialized, ballistic gas clouds. 

We can also use our HI velocity covering fraction analysis to learn about the origin of the HI gas in the IC region. To do this, we looked for a possible correlation between the HI velocity covering fraction and the satellite velocity distribution. In Figure~\ref{fig:fcov_velhist}, we show the velocity distribution of all satellite galaxies (all identified satellites with at least 100 cells and stellar 
mass $\geq$ $10^{8.5}$ $M_{\odot}$), as a function of the number of galaxies (orange leftmost y-axis of each panel) and their stellar mass (black left y-axis of each panel). The grey bars in Figure~\ref{fig:fcov_velhist} indicate the summed stellar mass of galaxies binned within velocity ranges of 100 $\rm km s^{-1}$. The orange histogram represents the number counts of all satellite galaxies in each velocity bin. Additionally, we overlay the velocity covering fraction (right y-axis of each panel) for log $\rm N_{HI}$ $\geq$ 18 $\rm cm^{-2}$ in blue (total and IC HI in solid and dashed lines, respectively). For most of the halos, we find that the velocity covering fraction for both the halo and IC regions follows the velocity distribution of satellite galaxies. This points towards a scenario where the IC HI gas originated from satellite galaxies.  We performed a similar analysis considering only the HI-rich galaxies (satellite having HI gas cells outside 10 $\rm R_{1/2*}$) and found similar correlation between the stellar mass distribution and HI velocity covering fraction, except that the stellar mass distribution is less centrally peaked (as we would expect for gas rich galaxies that have likely recently entered the cluster).

We show both the number distribution and the stellar mass distribution of satellites because both could effect the total HI mass brought into the cluster. The stellar mass distribution of satellite galaxies may even be more likely to predict where HI will be found as, for example, a single 10$^{10}$ M$_{\odot}$ galaxy is likely to fall into the cluster with more HI than two 10$^{9}$ M$_{\odot}$ galaxies. In Figure ~\ref{fig:fcov_velhist}, we note that besides the agreement with the number distribution of satellites, the HI velocity covering fraction also correlates with peaks in the velocity distribution of the stellar masses of all satellite galaxies (grey histogram). Indeed, in Halos 3 and 7, where the satellite stellar mass distribution shows multiple peaks, the HI velocity distribution seems to follow the stellar mass more closely than the number of satellites. This general agreement between the satellite and IC HI velocity distributions suggests that IC HI originates from satellite galaxies, and the possibly stronger relation between the satellite stellar mass and HI velocity gives a hint that the IC HI may even originate from more massive galaxies (although we stress that this does not necessarily imply the gas is stripped from massive disks but may fall in as part of the galaxy's circumgalactic gas). 

Although a detailed investigation is outside the scope of this paper, we note that it is expected that some processes occurring in clusters affect the gas dynamics, and therefore some offset between the HI and satellite velocity distributions is unsurprising.  For example, a past merger can cause long-lived motions in the ICM \citep{Iraj2022}, or black hole activity could affect gas dynamics \citep{Weinberger2017}.

\section{Discussion}\label{sec:discussion}

In this section, we discuss our results and compare the HI covering fraction to available observations and other cosmological simulation studies. 

\subsection{Comparison to other cosmological simulations}\label{sec:dis_sims}

We begin by comparing our measured cumulative HI covering fraction with the available HI covering fraction from the studies of \cite{nelson2020}. They measured the abundance of cold gas in TNG50 halos for massive halos with mass $>$ \pow{11} $\rm M_{\odot}$ at intermediate redshift z $\sim$ 0.5. Although we have only 7 TNG50 halos, and apart from the evolution of the halos from redshift z$\sim$0.5 to 0, the simulations and HI model we used are the same as \cite{nelson2020}; therefore, the HI covering fraction should be of the same order. Our measured HI covering fraction for \NH {} agrees well with the \cite{nelson2020} measured values. For a column density of \NH $>$ \pow{17} \cm, we find a covering fraction around 70 $\pm$ 15 \% at  10 kpc, dropping to 30 $\pm$ 15\% at 100 kpc, and for \NH $>$ \pow{20} \cm, at 100 kpc, the covering fraction is roughly 10\%, similar to the findings of \cite{nelson2020}. \cite{rahmati2015} used the EAGLE simulation to study the HI distribution around high-redshift massive galaxies. They found a strong evolutionary trend in the HI covering fraction within the virial radius with redshift. For an averaged HI column density in between \pow{17.3} $<$ \NH/\cm $<$ \pow{21}, the HI covering fraction drops from 70 \% at z = 4 to 10 \% at z =1. The HI content of galaxies in the EAGLE cosmological simulations was also investigated \citep{2016MNRAS.461.2630M, 2017MNRAS.464.4204C}, finding that the highest resolution simulations reproduced the HI masses of galaxies as well as their clustering. In addition, in dense group and cluster environments, they found that ram pressure stripping was the primary HI mass removal process but that galaxy interactions also played a role.

Studying the spatial distribution and ionisation state of cold gas in the CGM, \cite{Fareman2023} performed semi-analytical modelling of cold gas in the CGM of low-redshift star-forming galaxies. Assuming that cold clouds in the CGM are in local pressure equilibrium with the warm/hot phase, they reported that cold gas can be found out to 0.6 $\rm R_{\rm vir}$ or beyond. Although we examine more massive halos ($10^{13.5}$ $\leq$ $\rm M_{200}$ $\leq$ $10^{14}$ $M_{\odot}$) compared to \cite{Fareman2023}, we also find that the CGM of Fornax-like halos normally shows a spatially extended distribution of cold gas clouds out to more than 0.5 $\rm R_{\rm vir}$.

Previously, \cite{voort2019} have shown that standard mass refinement and a high spatial resolution of a few kpc scale can significantly change the inferred HI column density. Studying zoom-in simulations of a Milky-way mass galaxy within the virial radius, they found that the HI covering fraction of \NH $\leq$ \pow{19} \cm\ at 150 kpc is almost doubled from 18\% to 30\% when increasing the spatial resolution of the CGM. Although the simulation setup of \cite{voort2019} and TNG50 is different, we compare our findings to theirs based on the similar resolutions ($\sim$1 kpc). 
We find that the cumulative HI covering fraction within 150 kpc is around 25\%, quite close to the \cite{voort2019} result.   

In addition to the \cite{voort2019} work, several papers have focused on studying CGM properties using higher spatial and mass resolution simulations. For example, the FOGGIE group \citep{FOGGIE2019} has used the cosmological code Enzo \cite{Enzo} to carry out a set of simulations with high resolution in the CGM, finding that a great deal of small-scale structure emerged in the multiphase gas, producing many more small clouds.  However, they found that the HI covering fractions did not change significantly. Similarly, \cite{Hummels2019} studied the simulated CGM with an enhanced halo resolution technique in the TEMPEST simulation, again based on ENZO cosmological zoom simulations. They found that increasing the spatial resolution resulted in increasing cool gas content in the CGM. With an enhanced spatial resolution in the CGM, they found that observed HI content and column density increases in the CGM. In a similar vein \cite{Suresh2019} explored the CGM of a $10^{12}$ $M_{\odot}$ mass halo with a super-Lagrangian zoom in method, reaching up to $\sim$ 95 pc resolution in CGM. They reported that enhanced resolution results in an increased amount of cold gas in the CGM and this increase in the cold gas also results in a small increase in the HI covering fraction. For a column density of \NH $\geq$ \pow{19} \cm, they found a HI covering fraction value of around 10 $\%$ at one virial radius, which agrees well with our measured covering fraction of 12 $\%$ at one virial radius. More recently, \cite{Ramesh2024} re-simulated a sample of Milky Way galaxies at z $\sim$ 0 from TNG50 simulation with the super Lagrangian refinement method. Going down to a scale of 75 pc, they also reported that the abundance of cold gas clouds increases with enhanced resolution but did not find a large change in the covering fraction.

\subsection{Detection of HI clouds in the Intracluster (IC) region through observational work}\label{sec:dis_obs}

For all the TNG50 Fornax-like halos, we found that the HI clouds can be found beyond 0.5 $\rm R_{vir}$, which corresponds to an average physical scale of around 350 kpc. Although only a few, there are observational studies showing the existence of remote HI clouds associated with galaxies. An important example is the detection of HI clouds in the inter-galactic medium of the galaxy group HCG44, where the HI clouds extend to more than $\sim$ 300  kpc \citep{serra2013}. Another example is the case of NGC\,4532, in the Virgo cluster, where the HI tail of the galaxy with some discrete clouds extends to 500 kpc and constitutes around 10$\%$ of the total HI mass \citep{koopmann2007}. 

A number of observational properties of the multiphase nature of the CGM in groups and clusters have been explored through absorption lines of background quasars with the HST-Cosmic Origins Spectrograph (COS). Studying a sample of low redshift luminous red galaxies (LRG) with metal absorption lines such MgII, CIII, and SiIII, \cite{Werk2013} measured the occurrence of cool metal enriched CGM and reported a MgII ion covering fraction (down to very low column densities) of 0.5 within 160 kpc radius. Using a sample of 16 LRG at z $\sim$ 0.4 observed with the HST/COS, \citep{chen2018, zahedy2019} found a high HI covering fraction for column density \NH $>$ \pow{17.2} \cm of about 0.44 within 160 kpc impact parameter. In comparison to this, we measured an HI covering fraction of around 0.75 for column density \NH $>$ \pow{17.2} \cm at 160 kpc. Studying the CGM of a sample of 21 massive galaxies at z $\sim$ 0.5, \cite{Berg2019} measured an HI covering fraction of column density for \NH $>$ \pow{17.2} \cm within the virial radius of 15 $\%$, which closely agrees with our average measured HI covering fraction of 12 \%.  Finally, \cite{2015MNRAS.453.4051E} compared HI covering fractions within the virial radius in Virgo-like clusters between simulations and observations, finding values consistent with those found here.

Most recently, using the MeerKAT observations of the Fornax\,A subgroup, \cite{kleiner2021} reported the detection of HI clouds at $\sim$220 kpc from NGC\,1316, the central galaxy of Fornax\,A. Another study done with the MeerkAT telescope have detected a large extended HI cloud, extending $\sim$ 400 kpc in proximity to a large galaxy group at a redshift of z $\sim$ 0.03 \citep{Jozsa2022}. Although observational detections of HI clouds in the IC regions around massive galaxies are few and rare, our and other simulation work like \cite{rahmati2015, voort2019, nelson2020} strongly suggest the existence of dense small HI clumps within the ICM. 
For our work we find that within the IC region, HI tends to have a 
column density log $\rm N_{HI}$ $\sim$ 19 $\rm cm^{-2}$ or less, and current observations are mostly not that sensitive yet. A strong test of cosmological simulations will be to compare our predicted HI covering fractions with MFS observations. If simulations overpredict the covering fraction of cold gas, some combination of i) excess cold gas removal from satellites, ii) excess cold gas added to the ICM from feedback or filamentary accretion, and iii) suppressed heating of cold gas in the ICM are likely at play. On the other hand, if simulations underpredict the HI covering fraction, some combinations of these effects are leading to too little cold gas. 

\begin{figure}
\epsscale{1.2}
\plotone{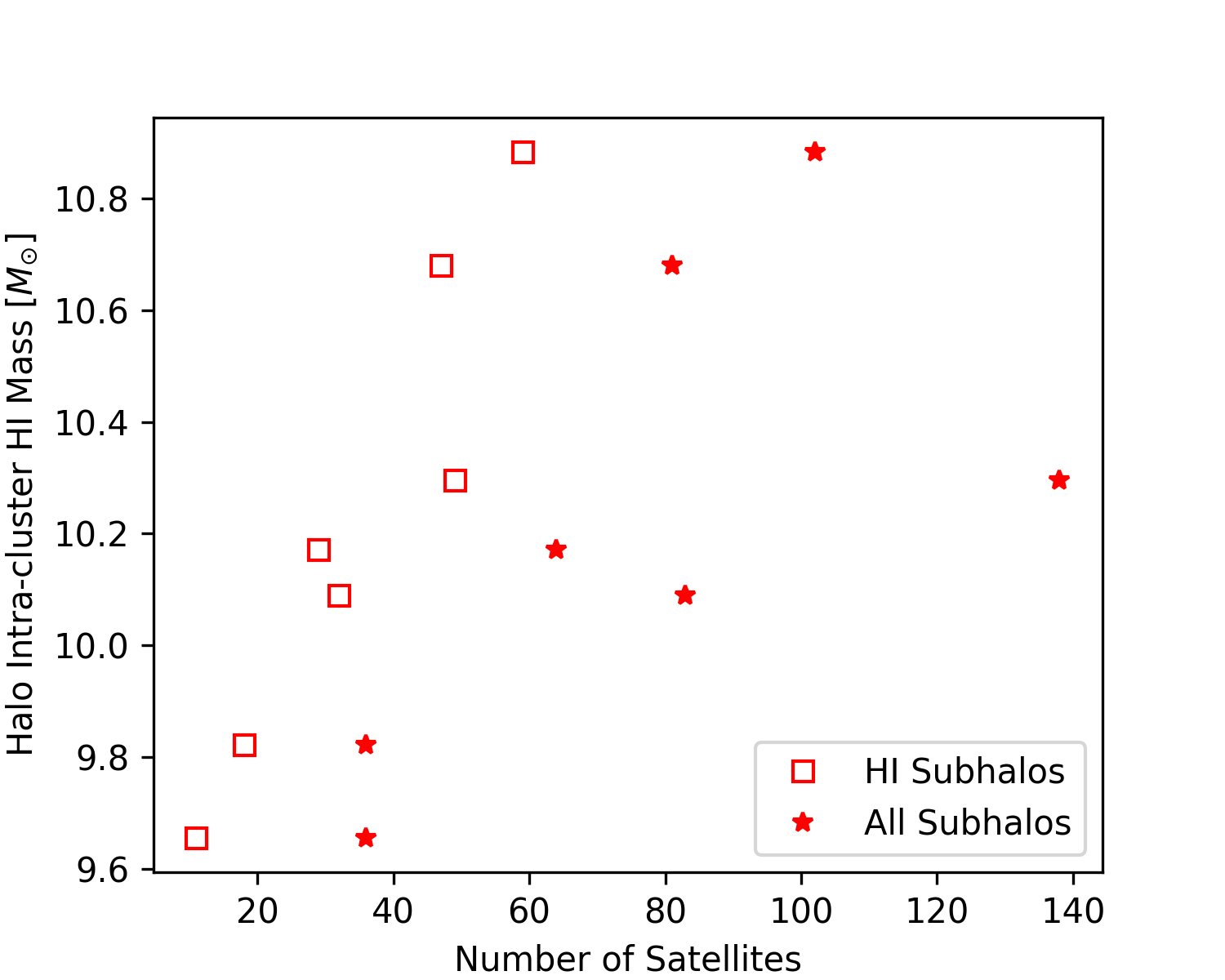}
\caption{HI intra-cluster (IC) mass as a function of number of satellites for our seven Fornax-like halos.}
\label{fig:HI_vs_satellite}
\end{figure}

\subsection{Possible Origin Scenario} \label{dis:origin}

A detailed study of the origin or production of the large amount of HI gas in TNG50 Fornax-like halos is beyond the scope of this work. However, we speculate here on the possible source of the HI gas we do find. \cite{nelson2020} studied the cold gas distribution in TNG50 massive halos at intermediate redshift z $\sim$ 0.5. Using Lagrangian tracer analysis, they argued that cold gas in TNG50 halos is related to gas that is removed from the halos in infalling satellites. These gas clouds can later stimulate the cooling process leading to a significant amount of cold gas. 

Most recently, using the TNG50 simulation \cite{Ramesh2024} studied the cold gas clouds in the CGM of Milky-Way like galaxies. They reported that  these high density gas clouds show clustering behaviour and this over-density increases around satellite galaxies, suggesting that a fraction of these clouds originate from ram-pressure stripping. This suggests that, not only for massive halos (like our Fornax-like halos), but even for smaller galaxies, the satellite gas stripping scenario can be important in producing HI clouds in the CGM. We however do not discount the possibility that a fraction of these clouds, whether around the satellite galaxies or isolated, may originate from the in-situ condensation of hot halo CGM gas in the cluster environment, or may be associated with outflows from the central galaxy \citep{Fraternali2006}. Studying the formation mechanism of high velocity clouds around the Milky-Way like disk galaxies, \cite{Binner2009, Fraternali2015} have suggested that condensation of hot CGM gas can produce the HI clouds. It will be interesting to learn which is the effective mechanism for producing these HI clouds in the CGM. Possibly all the mechanisms: a) satellite stripping, b) thermal instability in CGM, or c) the feedback from the galaxy play roles in the formation of these clouds and could be explored by characterising these cloud properties in phase space (spatial and velocity) such as was done in \cite{Ramesh2024}.

In addition to the analysis in Section \ref{sec:vel_covering_frac}, which indicates 
that the IC velocity covering fraction is associated with the satellite galaxies' velocity distribution, we checked if we could see any correlation between the satellite galaxies' number with the total IC HI mass. Figure~\ref{fig:HI_vs_satellite} shows the halo intracluster HI mass as function of satellite galaxy number. In Figure~\ref{fig:HI_vs_satellite}, red stars denote all of the satellite galaxies and open squares mark the satellites having HI gas cells outside their 10 $\times$ $\rm R_{1/2*}$. With only 7 halos, its hard to quantify any relation, but we find that the halo IC HI mass increases with increasing number of satellite galaxies. The IC HI mass correlates more steeply with satellites having HI and shows less scatter. This further suggests that the HI mass in the IC regions of Fornax-like clusters could be associated with the stripped HI gas from satellite galaxies, similar to the findings of \cite{nelson2020}. We find a similar level of correlation between the total stellar mass in satellites and the halo IC HI mass.

\section{Summary and Conclusion}\label{sec:summary}

In this paper, using the publicly available TNG50 simulation data, we have studied the distribution of HI gas in halos similar to the Fornax galaxy cluster. Adopting the MeerKAT Fornax survey (MFS) observational conditions, we have measured the HI covering fraction of the halos with a mass of $10^{13.5}$ $<=$ $\rm M_{200}$ $<=$ \pow{14} M$_{\odot}$. The following points summarise our findings and conclusions:

\begin{enumerate}
   
\item Atomic hydrogen in TNG50 Fornax-like halos shows a wide spatial distribution, appearing as clouds and filamentary structures (Figures \ref{fig:HIzoominmap} and \ref{fig:HImaphalo}). HI is non-uniformly distributed and extends in patches well beyond 0.5 virial radii of the central galaxy. On a physical scale, this corresponds to $\sim$ 350 kpc.

\item  Using our HI covering fraction measurements, we find that individual Fornax-like halos in TNG50 show a wide scatter in the measured HI covering fraction ranging from 3\% to 15\% at 1 $\rm R_{vir}$ (Figure \ref{fig:scatter}). We predict the upcoming MFS should observe a total HI covering fraction of $\sim$ 25\%  at 0.5 virial radii and $\sim$ 12\%  at 1 $\rm R_{vir}$ (Figure \ref{fig:cov_fraction_vs_pixelsize}) at \NH$ \geq$ \pow{18} \cm (spatial resolution $\sim$ 10 kpc). For intracluster regions, this values drops to $\sim$ 20\%  at 0.5 virial radii and $\sim$ 9\%  at 1 $\rm R_{vir}$.

\item  Intracluster (IC) regions (i.e. more than 10 stellar half-mass radii from identified galaxies) in Fornax-like halos hold a substantial fraction of the HI. When using the \NH\ $\geq$ \pow{18} \cm\ contour, the IC HI covering fraction at 1 $\rm R_{vir}$ (spatial resolution $\sim$ 10 kpc) corresponds to around 75\% of the total HI covering fraction (Figure \ref{fig:scatter}).

\item The HI velocity covering fraction for the Fornax-like halos (both in total and in the IC regions only) 
shows a broad velocity distribution that is not generally Gaussian, indicating that HI is not virialized in the halos (Figure \ref{fig:fcov_velocity}). The HI velocity covering fraction for both halo and IC largely follows the velocity distribution of  satellite galaxies, suggesting that IC HI is associated with satellite galaxies (Figure \ref{fig:fcov_velhist}).

\item  We find that halo HI intracluster mass increases with increasing number of satellite galaxies and shows an even stronger correlation with the satellites having HI presence in their outskirts (Figure \ref{fig:HI_vs_satellite}). This also suggests that HI in the IC regions is associated with the stripped gas of satellite galaxies, similar to the results of \cite{nelson2020}.

\end{enumerate}

With this work, we have demonstrated, based on TNG50 simulation data, that HI cold gas is predicted to co-exist and survive in the hot intracluster medium for Fornax-like clusters. Based on HI maps of Fornax-likes halos in TNG50, we expect MFS to find extended HI well beyond the satellites in the halo, but should generally follow the large-scale satellite distribution, both on the sky and in velocity space.  
This is also reflected in the asymmetry of the HI in velocity space - while the average distribution of the seven halos studied is symmetric, any individual halo can show strong asymmetries. We plan to perform a future follow-up study to pinpoint the origin of these HI clouds, whether they are possibly stripped or formed in situ in the cluster environment. It will be illuminating to see what MFS will observe within the Fornax cluster. With its higher sensitivity, there remains a good chance that the MeerKAT telescope can provide observational support and constraints for current and future simulation work on the multiphase nature of halo gas.

\section{ACKNOWLEDGEMENTS}
A.C. would like to thank Abhijeet Anand and Roland Sazacks for helpful suggestions and discussions regarding this work. A.C. acknowledges the financial and computing support from Simons Foundation and thanks the Flatiron Institute pre-doctoral fellowship program through which this research work was carried out.  
GLB acknowledges support from the NSF (AST-2108470, ACCESS), a NASA TCAN award, and the Simons Foundation through the Learning the Universe Collaboration.

\appendix

\section{TNG50 halos maps at 10 kpc pixel size}\label{appen1}

In our paper we use pixels that are 2 kpc on a side in order to match the MFS resolution at HI column densities of about 10$^{19}$ cm$^{-2}$.  However, the resolution at HI column densities of 10$^{18}$ cm$^{-2}$ is about 10 kpc.  In Figure \ref{fig:cov_fraction_vs_pixelsize} we show how the covering fraction at this column density increases with pixel size, and in this Appendix Figure \ref{fig:HImap_appen}, we show HI maps made with 10 kpc pixels so the reader can directly compare with the higher resolution maps shown in Figures \ref{fig:HImaphalo} and \ref{fig:HIzoominmap}.

\begin{figure*}[h]
\epsscale{0.70}
\plotone{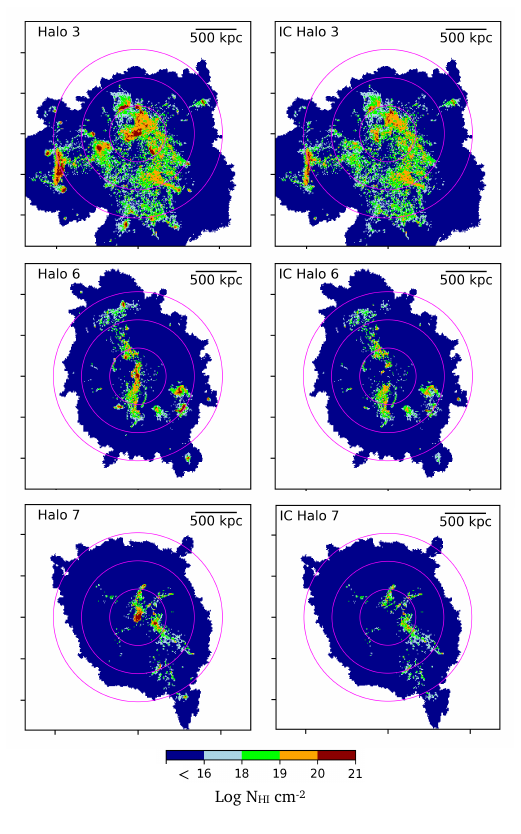}
\caption{Example of HI maps made with 10 kpc pixel scale, showing HI distribution in the TNG50 halo 3, 6 and 7 at redshift z=0, colour-coded with the HI column density. These maps are shown projected along the (arbitrarily chosen) z-axis. The three pink circles indicate the viral radius of the halos marked at 0.5, 1.0 and 1.5 times $\rm R_{\rm vir}$. Dark blue color in the maps indicate the HI column density lower than \NH= \pow{16} \cm. {\it Left:} Full halo HI distribution, {\it right:} HI distribution in the intracluster (IC) region.}
\label{fig:HImap_appen}
\end{figure*}

\section{TNG50 halos maps at out to 3 $\rm R_{\rm vir}$}\label{appen2}

For our study we considered only the gas cells gravitationally bound to a halo (Sec. \ref{sec:analysis}), which generally includes cells out to an average of 1 to 1.5 $\rm R_{\rm vir}$. Here in this Appendix Figure \ref{fig:HImap_appen_3rv}, we show TNG50 Fornax-like halos maps out to 3 $\rm R_{\rm vir}$, demonstrating that there are several infalling satellite galaxies, particularly for Halos 1, 2 and 4. Their contribution to the HI covering profile are minimal. These maps also confirm that we do not detect any signature of diffuse HI (not closely associated with satellite galaxies) in the outer IGM of halos).

\begin{figure*}
\epsscale{1.2}
\plotone{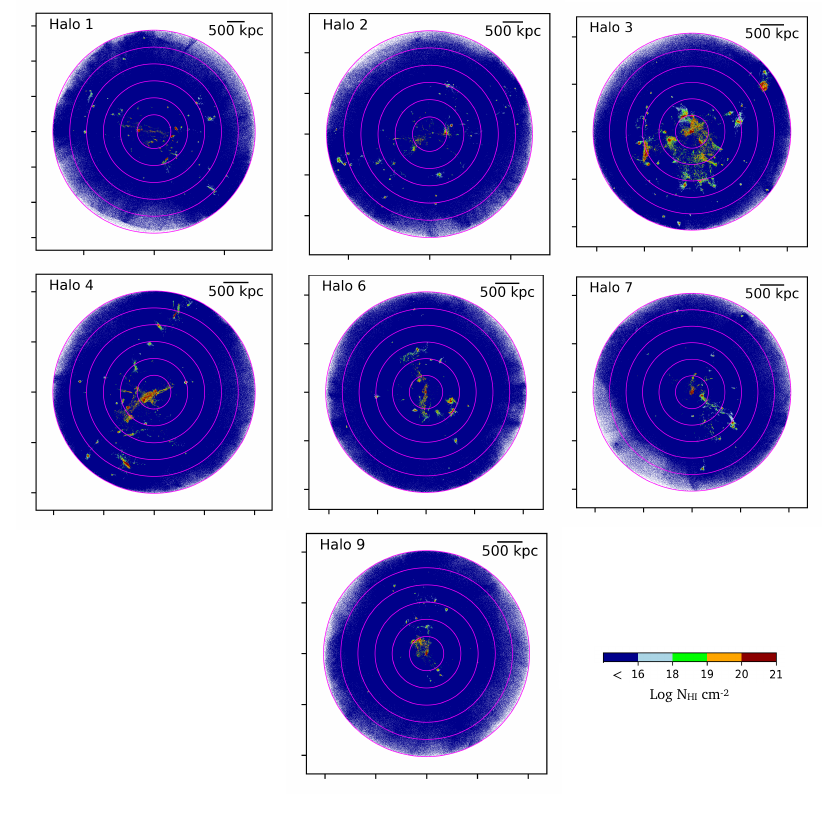}
\caption{HI maps made with 2 kpc pixel scale, showing HI distribution in for all the TNG50 Fornax-like halos at redshift z=0, colour-coded with the HI column density out to 3 $\rm R_{\rm vir}$. These maps are shown projected along the (arbitrarily chosen) z-axis. The six pink circles indicate the viral radius of the halos marked at 0.5, 1.0, 1.5, 2.0, 2.5, 3.0 times $\rm R_{\rm vir}$. Dark blue color in the maps indicate the HI column density lower than \NH= \pow{16} \cm. }
\label{fig:HImap_appen_3rv}
\end{figure*}

\bibliography{tng50mfs_avi}
\bibliographystyle{aasjournal}

\end{document}

%% file: tng50_fnal_halos.tex
1 & 13.97 & 9.59 & 10.80 & 384.24 & 446.20\\
2 & 13.81 & 8.46 & 10.37 & 373.80 & 402.12\\
3 & 13.54 & 6.92 & 11.24 & 306.22 & 321.83\\
4 & 13.50 & 6.71 & 11.07 & 204.56 & 345.08\\
6 & 13.54 & 6.88 & 10.67 & 279.20 & 320.58\\
7 & 13.52 & 6.80 & 10.38 & 348.00 & 331.76\\
9 & 13.51 & 6.75 & 10.49 & 254.10 & 325.88\\